\documentclass[twocolumn,prl,showpacs,preprintnumbers,amssymb]{revtex4-1}

\usepackage{epsfig}
\usepackage{dcolumn}
\usepackage{bm}

\usepackage{amsmath}
\usepackage{amssymb}
\usepackage{latexsym}
\usepackage{epsfig}
\usepackage{color}
\usepackage{mathtools}
\usepackage{youngtab}
\usepackage{bigints}
\usepackage{sidecap}
\usepackage{slashed}

\def\IC{\bf C}
\def\IZ{\bf Z}

\def\z2z2{$\IC^3/(\IZ_2\times\IZ_2)$}

\def\id{{\bf 1}}

\def\cG{\cal G}

\def\cp{\mbox{\bbbold C}\mbox{\bbbold P}}

\def\a{\alpha}

\def\b{\beta}

\def\d{\delta}\def\D{\Delta}

\def\k{\kappa}
\def\l{\lambda}

\def\p{\pi}

\def\s{\sigma}

\def\th{\theta}

\def\beq{\begin{equation}}\def\eeq{\end{equation}}
\def\beqa{\begin{eqnarray}}\def\eeqa{\end{eqnarray}}
\def\barr{\begin{array}}\def\earr{\end{array}}

\def\wt{\widetilde}

\def\ds {{\del \hspace{-6.4pt} \slash}\;}

 \let\br=\bigr

\def\bd{\begin{document}}
\def\ed{\end{document}}
\def\ba{\begin{array}}
\def\ea{\end{array}}
\def\bea{\begin{eqnarray}}
\def\eea{\end{eqnarray}}
\def\ft#1#2{{\textstyle{{\scriptstyle #1}\over {\scriptstyle #2}}}}
\def\fft#1#2{{#1 \over #2}}
\newcommand{\be}{\begin{equation}}
\newcommand{\ee}{\end{equation}}
\newcommand{\eq}[1]{(\ref{#1})}
\def\eqs#1#2{(\ref{#1}-\ref{#2})}
\def\det{{\rm det\,}}
\def\tr{{\rm tr}}
\newcommand{\ho}[1]{$\, ^{#1}$}
\newcommand{\hoch}[1]{$\, ^{#1}$}
\def\ra{\rightarrow}

\def\Xh{\hat{X}}
\def\ah{\hat{a}}
\def\xh{\hat{x}}
\def\yh{\hat{y}}
\def\ph{\hat{p}}
\def\G{{\cal G}}
\def\Dth{{\Delta_\th}}

\def\bk{{\bf k}}
\def\bx{{\bf x}}
\def\br{{\bf r}}
\def\tr{{\rm tr \,}}
\def\Tr{{\rm Tr \,}}
\def\diag{{\rm diag \,}}
\def\tg{{\rm tg \,}}
\def\NPB#1#2#3{Nucl. Phys. B {\bf #1} (19#2) #3}
\def\PLB#1#2#3{Phys. Lett. B {\bf #1} (19#2) #3}
\def\PLBold#1#2#3{Phys. Lett. {#1B} (19#2) #3}
\def\PRD#1#2#3{Phys. Rev. D {\bf #1} (19#2) #3}
\def\PRL#1#2#3{Phys. Rev. Lett. {\bf #1} (19#2) #3}
\def\PRT#1#2#3{Phys. Rep. {\bf #1} C (19#2) #3}
\def\MODA#1#2#3{Mod. Phys. Lett.  {\bf #1} (19#2) #3}
\def\ov{\overline}

\def\preal{{\rm Re\,}}
\def\pim{{\rm Im\,}}

\def\ds{\displaystyle}
\def\yzero{\smash{\hbox{$y\kern-4pt\raise1pt\hbox{${}^\circ$}$}}}
\def\p{\partial}
\def\a{\alpha}
\def\b{\beta}
\def\g{\gamma}
\def\d{\delta}
\def\beq{\begin{equation}}
\def\eeq{\end{equation}}
\def\beqa{\begin{eqnarray}}
\def\eeqa{\end{eqnarray}}
\def\Om{\Omega}
\def\om{\omega}
\def\th{\theta}
\def\vt{\vartheta}
\def\vphi{\varphi}
\def\-{\hphantom{-}}
\def\ov{\overline}
\def\s2{\frac{1}{\sqrt2}}
\def\wh{\widehat}
\def\wt{\widetilde}
\def\oh{\frac{1}{2}}
\def\tr{{\rm tr \,}}
\def\Tr{{\rm Tr \,}}
\def\diag{{\rm diag \,}}
\def\vac{|0 \rangle}
\def\vm{\relax{n_{\text{v}}}}
\def\cc{{\cal C}}
\def\ck{{\cal K}}
\def\ci{{\cal I}}
\def\cu{{\cal U}}
\def\cG{{\cal G}}
\def\cn{{\cal N}}
\def\cam{{\cal M}}
\def\cp{{\cal P}}
\def\ct{{\cal T}}
\def\cv{{\cal V}}
\def\cz{{\cal Z}}
\def\ch{{\cal H}}
\def\cf{{\cal F}}
\def\tv{\tilde v}
\def\Dsl{\,\raise.15ex\hbox{/}\mkern-13.5mu D} 
\def\IZ{Z\kern-.4em  Z}
\def\id{{\rm 1}}

\def\ti{\times}
\def\til{\tilde}
\def\eps{\epsilon}
\def\k{\kappa}
\def\A{\Arrowvert}
\def\cw{{\cal W}}
\def\G{\Gamma}
\def\car{{\cal R}}
\def\l{\lambda}
\def\raw{\rightarrow}
\def\Raw{\Rightarrow}
\def\inte{{\bf Z}}
\def\cpx{{\bf C}}
\def\real{{\bf R}}
\def\Lam{\Lambda}
\def\D{\Delta}
\def\cb{{\cal B}}
\def\ca{{\cal A}}
\def\Z{\mathbb{Z}}

\definecolor{mygr}{rgb}{0,0.6,0}
\definecolor{mygrey}{rgb}{0,0.1,0.2}
\definecolor{myblue}{rgb}{0,0.5,0.9}
\definecolor{myblue2}{rgb}{0,0.5,0.5}
\definecolor{myorange}{rgb}{1,0.5,0}
\definecolor{mypurple}{rgb}{0.6,0,1}
\definecolor{mygolden}{rgb}{1,0.8,0.2}
\definecolor{myblue3}{rgb}{0,0.1,0.9}
\definecolor{myblue4}{rgb}{0,0.6,0.6}

\begin{document}

\preprint{MAD-TH-17-08~ IFT-UAM/CSIC-18-69
}

\title{Strong Dynamics and Natural Inflation}

\author{Gary Shiu$^{1}$, and Wieland Staessens$^{2}$,}
\affiliation{\small\slshape  $^{1}$ Department of Physics, 1150 University Avenue, University of Wisconsin-Madison, Madison, WI 53706, USA \\
     $^{2}$ Instituto de F\'isica Te\'orica UAM-CSIC, Cantoblanco, 28049 Madrid, Spain
      }
\begin{abstract}
We continue our investigation of the 4d effective field theory for closed string axions in Type~II compactifications with D-branes. The inclusion of St\"uckelberg couplings for the axions requires the presence of chiral fermions at D-brane intersections, whose interactions at strong non-Abelian gauge coupling induce mass terms for the axions and scalar chiral condensate excitations, dubbed infladrons. The set-up allows for a realization of natural-like inflation with a closed string axion as inflaton and a flattened scalar potential due to the back-reaction of the more massive infladrons. We further point out that this large field inflationary model is not compromised by axionic wormhole corrections. 
\end{abstract}
\maketitle

\section{Introduction}
Axions or axion-like particles arise abundantly~\cite{Witten:1984dg,Choi:1985je,Svrcek:2006yi} when ten-dimensional superstring theories are compactified to a four-dimensional spacetime. These axions correspond to the Kaluza-Klein (KK) zero mode scalars of higher-dimensional $p$-form gauge potentials ${\cal C}_{(p)}$ under the KK-reduction. As such, they inherit their shift symmetries from the higher dimensional gauge invariance under (large) gauge transformations for ${\cal C}_{(p)}$. By introducing a basis of $p$-cycle $\gamma_i$, each stringy axion can be associated to a cycle according to,
\begin{equation}
c^i = \frac{1}{2\pi} \int_{\gamma_i} C_{(p)} \qquad i\in \{1, \ldots , b_p \},
\end{equation}
with the Betti-number $b_p$ giving the dimensionality of the homology group $H_p({\cal M}, \mathbb{R})$. 
As their perturbative interactions are fully constrained by their continuous shift symmetry, axions provide for lagrangians suitable to accommodate beyond the standard model physics, both in particle physics~\cite{Peccei-Quinn,Weinberg:1977ma,Wilczek:1977pj} as well as in cosmology~\cite{Turner:1989vc,Freese:1990rb}. For example, the shift symmetry is invoked in inflationary models to prevent a violation of the slow-roll conditions by uncontrollable perturbative corrections. But not all corrections to the axion scalar potential are fully under control, as the global shift symmetry is expected to be broken by quantum gravity effects~\cite{Abbott:1989jw}. Moreover, higher-dimensional Chern-Simons couplings for the gauge potentials ${\cal C}_{(p)}$ in string theory induce extra four-dimensional interactions which either break or gauge the global shift symmetry of their respective axion. It is therefore of utmost importance to fully understand the four-dimensional physics of stringy axions and investigate whether these models are further constrained by considerations coming from quantum gravity.
  
In recent years a lot of attention has been devoted to the {\it breaking} of the global axion shift symmetry to a discrete one following the coupling to a four-form~\cite{Dvali:2005an}. These axion-4-form couplings (which have been applied to inflation in \cite{Kaloper:2008fb}) appear naturally for closed string axions in Type II string theory compactifications with internal RR-fluxes and NS-fluxes~\cite{Bielleman:2015ina,Escobar:2015fda}, and represent a realization of the axion monodromy scenarios proposed in \cite{Marchesano:2014mla,Blumenhagen:2014gta}. In case this four-form represents the topological density of a non-Abelian gauge theory with field strength $G$, the breaking of the shift symmetry is rather seen as the consequence of non-perturbative axion couplings to gauge instantons:
\begin{equation}
{\cal S}_{\rm anom} = \int \frac{1}{8 \pi^2} \sum_i n_ic^i \Tr(G\wedge G).
\end{equation}
In Type II superstring theory such anomalous couplings arise by including in the compactification spacetime filling D$(p+3)$-branes wrapping $p$-cycles $\gamma^i$ along the internal directions, with $n^i \in \mathbb{Z}$ representing the topological wrapping numbers. String theory also allows for a different class of non-perturbative effects breaking the axion shift symmetry, namely a single Euclidean D$(p-1)$-brane wrapping the three-cycle~$\gamma^i$ yielding a non-perturbative correction to the superpotential:
\begin{equation}\label{Eq:NonPertSuper}
{\cal W}_{np} = A\, e^{- {\cal S}_{E} + i\, c^i}, \qquad {\cal S}_{E}  = {\rm Vol}(\gamma^i).
\end{equation} 
The coupling to E$(p-1)$-brane instantons breaks the continuous shift symmetry to a discrete one, $c^i \rightarrow c^i + 2\pi$, determining the topology of the closed string axion moduli space. There also exist E$(p-1)$-brane instantons wrapping various cycle $\gamma^i$ simultaneously, to which the linear axionic combination $p_i c^i$ couples. 

\begin{figure}[h]
\begin{center}
\vspace{-0.3in}\includegraphics[scale=0.25]{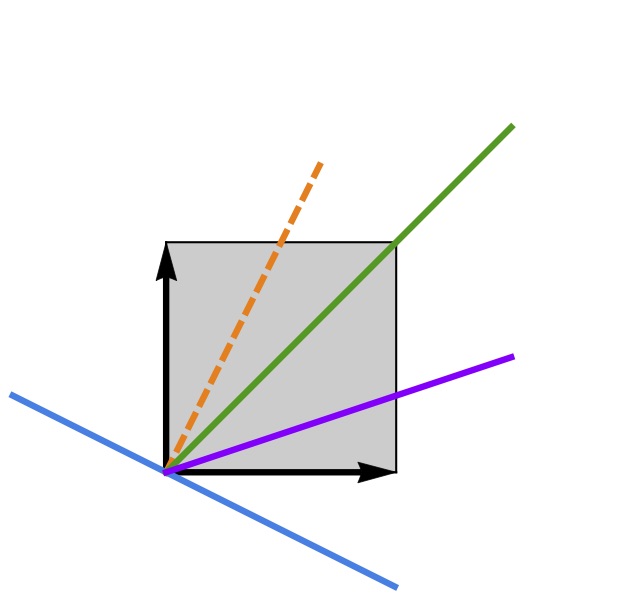}\begin{picture}(0,0) \put(-50,30){$c^1$}  \put(-118,95){$c^2$} \put(-95,115){\color{myorange}$(k^1,k^2)$} \put(-26,120){\color{mygr}$(n_1,n_2)$} \put(-55,0){\color{myblue}$(-k^2,k^1)$} \put(-25,60){\color{mypurple}$(p_1,p_2)$} \end{picture}
\end{center}
\caption{Closed axion moduli space with the $U(1)$ orbit under St\"uckelberg gauging (dashed orange line) and its orthogonal direction (blue line), the anomalous coupling to the non-Abelian gauge group (green) and the coupling to D-brane instantons (purple).  \label{Fig:AxionModuliSpace}}
\end{figure}
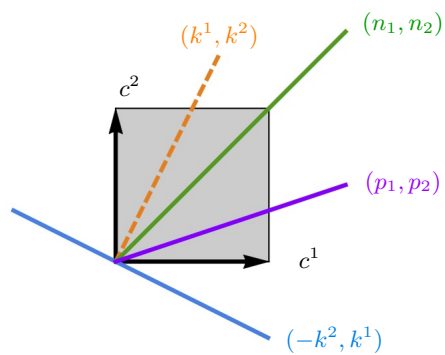

Alternatively, the axion shift symmetry can be {\it gauged} by virtue of a St\"uckelberg coupling to a $U(1)$ symmetry, which occurs when a D$(p+3)$-brane wraps the Poincar\'e dual $(6-p)$-cycle to the cycle $\gamma^i$ (combined with a flux-threaded $2$-cycle in case $p=4$). In this case, the axion acts as the longitudinal mode of the massive gauge boson living on the D-brane worldvolume. These St\"uckelberg charges are tied to the Green-Schwarz mechanism for D-branes which ensure anomaly cancelation in case of an anomalous $U(1)$ symmetry~\cite{Blumenhagen:2005mu}. Clearly, Type~II compactifications with multiple D-branes yield a rich four-dimensional effective field theory (EFT) for closed string axions, which already comes to light for a set-up consisting of two closed string axions $c^i$ both carrying a St\"uckelberg charge $k^i\neq 0$ under the same $U(1)$ gauge group with gauge boson $A$ supported by a D$(p+3)$-brane and both coupling anomalously to a non-Abelian gauge group $U(N)$ with field strength $G$ living on a stack of $N$ coincident D$(p+3)$-branes:
\begin{widetext}
\begin{align}   
{\cal S}^{\rm eff} =& \int \left[ - \frac{1}{2} \sum_{i=1}^2 {\cal G}_{ij} \big(dc^i - k^i A\big) \wedge\star_4 \big(dc^j - k^j A\big)  - \frac{1}{g_{U(1)}^2} \big|F\big|^2 - \frac{1}{g_{2}^2}  \Tr\big|G\big|^2 + \frac{1}{8 \pi^2} \left(\sum_{i=1}^2 n_i c^i \right)   \Tr(G\wedge G) \right. \nonumber \\
&\left.  \hspace{0.4in} + i \sum_{i=1}^{N_f} \ov \psi_L^i \slashed{D} \psi_L^i  + i \sum_{i=1}^{N_f} \ov \psi_R^i \slashed{D} \psi_R^i  \right], \label{Eq:FullStringModelEFTMS}
\end{align}
\end{widetext}
where the shorter notation $\big| C\big|^2=C\wedge \star_4 C$ is used for the $p$-form kinetic terms ($p\leq 4$) and $F$ denotes the field strength of the $U(1)$ gauge boson. Cancelling the mixed $U(1)-U(N)^2$ gauge anomaly by virtue of the Green-Schwarz mechanism  requires the presence of fermions $(\psi_L^i, \psi_R^i)$ chirally charged under the $U(1)$ group and transforming in a complex representation of $SU(N)$ (fundamental or anti-fundamental in practice), with the coupling to gravity and the gauge theories encoded in the Lorentz- and gauge-covariant derivative $\slashed{D}$. The matrix ${\cal G}_{ij}$ represents the metric on the axion moduli space, whose entries depend on the closed string metric moduli (the saxionic partners belonging to the same ${\cal N}=1$ 4d supermultiplet as the closed string axions) and are assumed to be stabilized. In this model, a linear axionic combination is eaten away by the $U(1)$ gauge boson, while the orthogonal direction $\xi$ is subjected to a potential resulting from the anomalous coupling to the non-Abelian gauge group. By properly identifying the eigenbases of the axionic mixing effects in kinetic and potential terms, one deduces~\cite{Shiu:2015uva,Shiu:2015xda} the effective decay constant $f_\xi$ for axion $\xi$ describing the coupling strength to $\Tr(G\wedge G)$, which depends on the integer numbers $(n_i, k^i)$ and the continuous geometric moduli. The intricate expression for $f_\xi$ allows for regions in the moduli space where the decay constant is enhanced or suppressed, suggesting a much broader window~\cite{Shiu:2015uva,CPSS2} than traditionally assumed in the literature:
\begin{equation}
10^{13} \text{ GeV} \lesssim f_\xi \lesssim 10^{19} \text{ GeV},
\end{equation} 
for  entries $\sqrt{{\cal G}_{ij}}$ of the order ${\cal O}(10^{16})$ GeV. Figure~\ref{Fig:AxionModuliSpace} offers a schematic view of the axion moduli space, including all possible mixing effects and couplings for the closed string axions in the presence of D-branes.

\section{Strong Dynamics and Infladrons}
Naively, the $\theta$-vacuum structure of the non-Abelian gauge theory in the strongly coupled regime is expected to induce a periodic scalar potential~\cite{Witten:1998uka} for inflaton candidate~$\xi$. But the presence of chiral fermions requires more precision to deduce the intricacies of the full vacuum, due to the global remnant of the chiral $U(1)$ symmetry below the St\"uckelberg scale $M_{st}= g_{U(1)} M_{\text{string}}$:
\begin{equation}\label{Eq:ChU1Dirac}
\begin{array}{rcl}
\psi^i &\rightarrow& e^{i\, q_+ \alpha}  e^{i\, q_- \alpha \gamma^5}\, \psi^i, \\
 \ov \psi^i &\rightarrow&  \ov\psi^i\, e^{-i\, q_+\alpha}  e^{i\, q_- \alpha \gamma^5} ,
 \end{array}
\end{equation}
for generation-independent charges $q_\pm = \frac{q_R \pm q_L}{2}$~\cite{comment0}. The axial-vector part of the associated current is not conserved at one-loop due to the chiral anomaly~\cite{Fujikawa:1979ay} and this non-invariance translates into effective interactions between chiral fermions in the instanton background~\cite{'tHooft:1976up,Callan:1977gz}:
\begin{equation}\label{Eq:GaugeInstantonMassTerm}
{\cal L}_{IC} = \frac{e^{- \frac{8\pi^2}{g_2^2} }}{g_2^{4N}}\left[ {\cal C}_{N_f}  \, e^{i \theta}  \det \left(\ov \psi^i (1 + \gamma^5) \psi^j \right) + h.c. \right],
\end{equation}
which break the global $U(1)$ symmetry to a discrete $\Z_{|q_-|N_f}$ symmetry. In this expression the determinant is evaluated over the number of generations $N_f$, $g_2$ corresponds to the renormalised gauge coupling of the non-Abelian gauge group, and ${\cal C}_{N_f}$ is a dimensionful parameter whose mass dimension depends on the number $N_f$ of chiral fermions. The 't Hooft interactions (\ref{Eq:GaugeInstantonMassTerm}) play a non-negligible r\^ole near the strong coupling scale $\Lambda_s=M_{\text{string}} e^{-\frac{8 \pi^2}{\beta_0 {g_{2}^2 (M_{\rm string})}}}$, provided that the one-loop beta-function coefficient $\beta_0= \frac{11}{3} N - \frac{2}{3} N_f$ is positive~\cite{Comment1}. In case this condition is satisfied without a Banks-Zaks fixed point~\cite{Banks:1981nn,Comment2}, the non-Abelian gauge theory finds itself in the confining phase near energy scales of the order ${\cal O}(\Lambda_s)$ and the free vacuum of massless fermions is no longer the true ground state of the theory~\cite{Casher:1979vw}. Instead, the fermionic contribution to the vacuum is more appropriately described in terms of a fermionic condensate consisting of neutral fermion-antifermion bound states with a non-vanishing vacuum expectation value $\langle \ov \psi^i \psi^j\rangle \sim \Lambda_s^3$, which breaks the remnant $\Z_{|q_-|N_f}$ discrete symmetry of the global $U(1)$ further down to a $\Z_2$ symmetry. 
 
The 't Hooft operator (\ref{Eq:GaugeInstantonMassTerm}) is not the only effective interaction among the chiral fermions at strong coupling. Integrating out the massive gauge boson also leads to effective four-fermion interactions~\cite{Shiu:2015xda}:
\begin{equation}
{\cal L}_{4 \psi} = \frac{1}{2 M_{st}^2} \big|{\cal J}_{\psi}\big|^2,
\end{equation}
where the one-form current $J_\psi$ corresponds to the exchange term initially coupling to the $U(1)$ gauge potential. In local (flat) coordinates, the one-form current reads:
\begin{equation}
J^\mu_\psi = \sum_{i=1}^{N_f} \left( q_L \ov \psi_L^i \gamma^\mu \psi_L^i + q_R \ov \psi_R^i \gamma^\mu \psi_R^i \right). 
\end{equation}
The four-fermion interactions can be further simplified through Fierz-identities~\cite{CPSS2}:
\begin{align}
{\cal L}_{4 \psi} =&  \frac{q_L q_R}{2 M_{st}^2} \sum_{i,j=1}^{N_f}  \left[ (\ov\psi^i \psi^j) (\ov \psi^j \psi^i) - (\ov\psi^i \gamma^5 \psi^j) (\ov \psi^j \gamma^5 \psi^i)    \right] \nonumber \\
& + \frac{1}{2 M_{st}^2} \sum_{i,j=1}^{N_f} \left[ q_L^2 (\ov\psi^i_L \gamma^\mu \psi^j_L)(\ov\psi^j_L \gamma_\mu \psi^i_L)  \right. \label{Eq:StringTheoryNJL} \\
& \hspace{0.8in}\left.+ q_R^2 (\ov\psi^i_R \gamma^\mu \psi^j_R) (\ov\psi^j_R \gamma_\mu \psi^i_R)\right],  \nonumber 
\end{align}
and it is easy to verify that these generalized Nambu-Jona-Lasinio (N-JL) four-fermion couplings~\cite{Nambu:1961tp} remain invariant under the $U(1)$ transformation in Eq.~(\ref{Eq:ChU1Dirac})~\cite{Comment3} and an accidental global $SU(N_f)_L\times SU(N_f)_R$ symmetry.

Below the strong coupling scale $\Lambda_s$, the description of Eqs.~(\ref{Eq:FullStringModelEFTMS}),  (\ref{Eq:GaugeInstantonMassTerm}) and (\ref{Eq:StringTheoryNJL}) in terms of interacting fermions and gauge bosons has to be replaced with an EFT in terms of interacting bound states of fermions and gauge bosons, which we will dub {\it infladrons}. Choosing a non-linear $\sigma$-model description with $N_f=1$ for simplicity, we parametrise the spin-zero bound state $\Phi = (f+\sigma(x)) e^{i \frac{\eta(x)}{f}}$ with $\sigma$ and $\eta$ describing the excitations over the vacuum $\langle \Phi\rangle = f \sim \Lambda_s$. The effective lagrangian for the field $\Phi$ then reads:
\begin{equation}\label{Eq:EFTNLRen}
\begin{array}{rcl}
{\cal L}^{\rm NL}_{\rm EFT} &=& - \frac{1}{2} \partial_\mu \xi \partial^\mu \xi - \frac{1}{2} (\partial_\mu \Phi)^\dagger \partial^\mu \Phi + \mu^2 \Phi^\dagger \Phi - \frac{\lambda}{2} ( \Phi^\dagger \Phi)^2 \\
&&  - \Lambda_{cc}+ [ \Lambda_s^2\,  \kappa\, e^{i\theta} \det (\Phi) e^{ i  \frac{\xi}{f_{\xi}} } + \Lambda_s^2 M  \Phi     + h.c. ], 
\end{array}
\end{equation} 
where the parameters $\mu$ and $\lambda$ are such that they trigger spontaneous U(1) symmetry-breaking and $\Lambda_{cc}$ is the four-dimensional cosmological constant of the Minkowski/de Sitter spacetime arising from moduli stabilization and other vacuum energy contributions. $U(1)$ symmetry-breaking due to the gauge instanton background is encoded by the parameter $\kappa e^{i\theta}~\sim~\Lambda_s e^{-\frac{8\pi^2}{g_2^2(\Lambda_s) }}e^{i\theta}$, resulting from the dominant contributions in the instanton zero mode measure at the strong coupling scale $\Lambda_s$. In the $\theta$-vacuum, the fermionic zero-modes in the gauge instanton background contribute to the non-vanishing bilinear $\langle \ov\psi \psi \rangle$ for $N_f=1$~\cite{Callan:1977gz,CPSS2}, such that the N-JL interactions induce an effective mass $M$:
\begin{equation}\label{Eq:FermMassConfinement}
M= - \frac{q_L q_R}{M_{st}^2} \langle \ov \psi_L \psi_R \rangle ,
\end{equation}
whose explicit $U(1)$-breaking is captured by the second linear term in~$\Phi$. The anomalous coupling of the axion $\xi$ to the non-Abelian topological density results in a dynamical $\theta$-term coupling to the 't Hooft determinant. 
Hence, the presence of the gauge instanton and scalar condensate backgrounds is sufficient for $N_f=1$ to generate mass terms for all scalar excitations and lift the mass degeneracy between the CP-odd scalars. In the limit where $f \ll f_{\xi}$, the diagonalised axion mass matrix offers the following mass spectrum~\cite{CPSS2}:
\begin{equation}
\begin{array}{rcl}
m_-^2 &=& 2  \frac{ f M \kappa}{f_{\xi}^2 (M  + \kappa) } \Lambda_s^2 + {\cal O}\left(\frac{f^2}{f_{\xi}^2}\right),\\
m_+^2 &=&2 \frac{ M + \kappa}{f}  \Lambda_s  +  {\cal O}\left(\frac{f^2}{f_{\xi}^2}\right), \\
m_\sigma^2 &=& 4 f^2 \lambda + m_+^2,
\end{array}
\end{equation}
in terms of the corresponding axion eigenbasis, 
\begin{equation}
a_+ = \eta + {\cal O}\left(\frac{f}{f_{\xi}}\right) \xi, \quad a_- = - {\cal O}\left(\frac{f}{f_{\xi}}\right)  \eta + \xi.
\end{equation}
For a non-Abelian factor $N\geq 3$, and a non-Abelian gauge coupling $g_{2}^2 \gtrsim 0.62$ at a string scale $M_{\text{string}}~\sim~{\cal O}(10^{17}~-~10^{18} \text{GeV})$, one easily obtains a strong coupling scale $\Lambda_s~\sim~{\cal O}(10^{15}-10^{16}\text{GeV})$ so that the mass spectrum exhibits a mass splitting with the following pattern:
\begin{equation}
m_- \sim {\cal O}(10^{13} \text{GeV}) \ll m_+< m_\sigma \sim {\cal O}(\Lambda_s),
\end{equation}
which is the desired mass scale for the inflaton mass $m_-$.

\section{Inflation and Infladron Corrections}
To extract a slow-roll single field inflationary model out of the three-field model in Eq.~(\ref{Eq:EFTNLRen}), one requires a firm handle on the quantum corrections to the scalar potential. Allowing for non-renormalizable corrections to Eq.~(\ref{Eq:EFTNLRen}) leads to a set of higher order derivative and interactions terms that are compatible with the symmetries of the model, causality and locality, as prescribed by Weinberg's theorem~\cite{Weinberg:1978kz}. These terms are generically suppressed by powers of the UV-cutoff scale $\Lambda_s$ or $M_{St}$, and many receive an additional suppression of the order ${\cal O}(f^2/f_\xi^2)$, such that it is allowed to work strictly with the renormalizable terms in Eq.~(\ref{Eq:EFTNLRen}). Also the $n$-loop perturbative corrections to the scalar potential due to quartic self-couplings of the infladrons are fully under control, upon solving the Callan-Symanzik equation for the effective action~\cite{Coleman:1973jx}. The composite nature of the infladrons combined with loop corrections can yield~\cite{Hill:1991jc} a non-minimal coupling of the infladrons to gravity of the form $\frac{\varpi}{2}\Phi^\dagger \Phi R$, yet RG evolution drives $\varpi$ to the fixed point $\varpi = 0$ in the IR, such that non-minimal coupling can be neglected~\cite{CPSS2}. 

\begin{figure}[h]
\begin{center}
\hspace*{-0.2in}\includegraphics[scale=0.17]{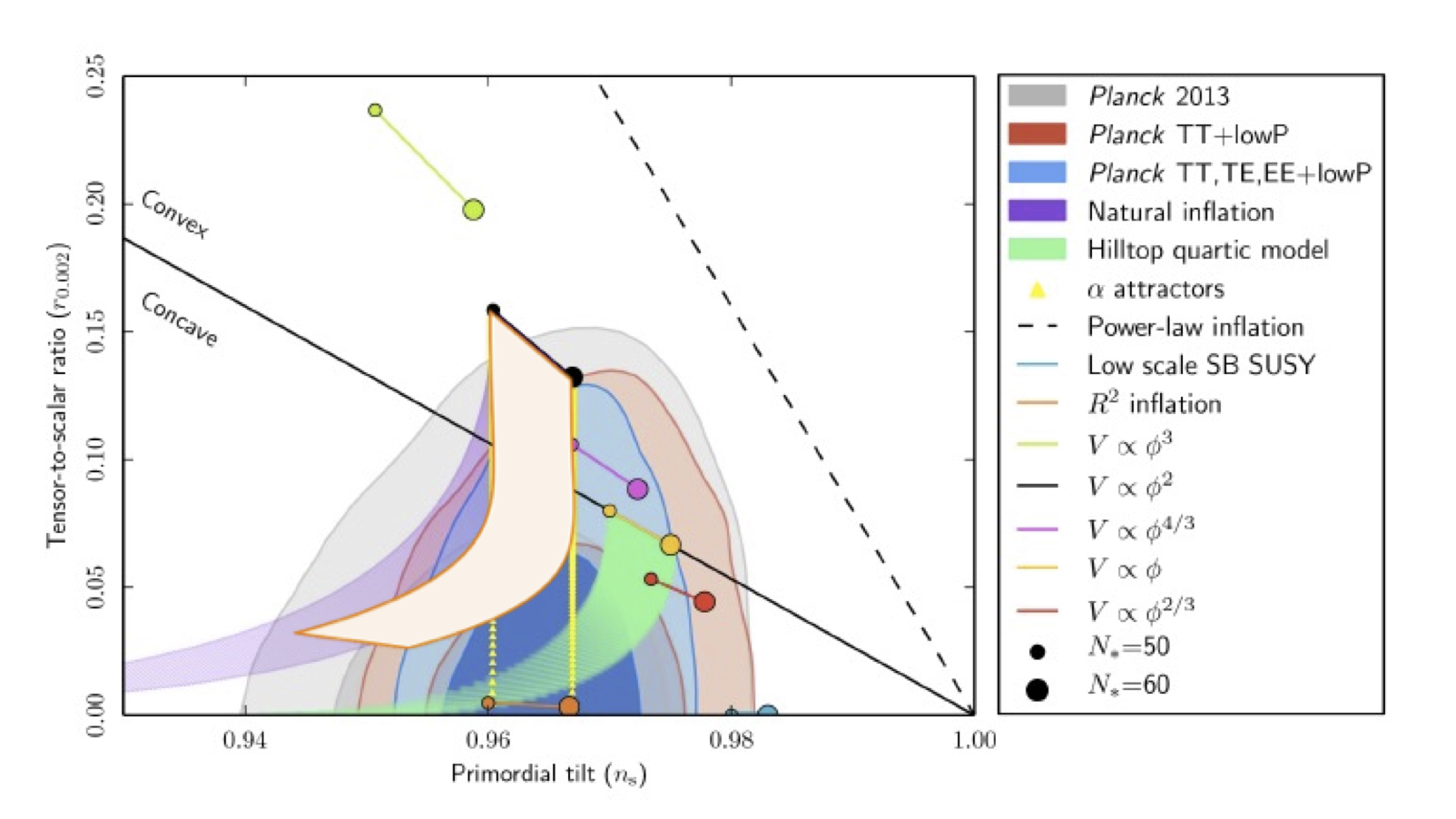}
\caption{$(n_s,r)$-plane for various inflationary models (taken from~\cite{Ade:2015lrj}) with the predictions for the backreacted natural inflation model represented by the orange strip, for parameter choice $\Lambda_s=\mu = 10 \lambda =\kappa = 2 M = 10^{16}$ GeV and $\log_{10} (f_\xi/M_{Pl})$ within the prior $[0.57,2.55]$. \label{Fig:nsrplanebackreaction}}
\end{center}
\end{figure}

A more worrisome effect is the back-reaction of the stabilized infladrons on the inflationary trajectory for the axion $\xi$, due to their displacement from their vacuum configuration during inflation. These displacements $\delta\sigma$ and $\delta\eta$ can be calculated as a function of the inflaton~$\xi$ from the vacuum constraint equations:
\begin{equation}\label{Eq:LinearBackreaction}
\delta \sigma (\xi) = \frac{2 M \Lambda_s}{m_\sigma^2}  \left( \cos \frac{\xi}{f_\xi}  - 1 \right), \qquad \delta \eta (\xi) = - \frac{\delta \sigma (\xi)}{f_\xi} \xi,
\end{equation}
and remain sufficiently small in case the Hubble scale $H_{\rm inf}$ during inflation is much smaller than $m_\sigma$, e.g. $m_\sigma\sim 10^{16}$ GeV for a scale $H_{\rm inf} \sim 10^{14}$ GeV. Taking into account the back-reaction of the stabilized infladrons leads to a backreacted scalar potential for~$\xi$, which is flattened at the maxima for a sufficiently large mass parameter $M<\kappa$. The flattening in the scalar potential can have visible repercussions for the cosmological observables, as shown explicitly in the $(n_s,r)$-plot of figure~\ref{Fig:nsrplanebackreaction}: the predicted results (orange strip) for the backreacted potential in the (Planck-data inspired) window $3.7 M_{Pl} \lesssim f_\xi \lesssim 350 M_{Pl}$~\cite{Ade:2015lrj} move towards the 95\% confidence region of the observational data for increasing $M$.  

\section{Wormholes and Weak Gravity}
Apart from field theory corrections, one equally has to worry about gravitational corrections and quantum gravity constraints invalidating the inflationary set-up. The $U(1)$ symmetry constrains the form of perturbative graviton loop-corrections, eliminating any perturbative danger to compromise the model. The greatest threat comes from axionic wormholes, which are supported by axion charges $w_\xi$ associated to $\xi$ and/or by axion charges~$w_\eta$ associated to the infladron $\eta$. Given that the EFT in Eq.~\eqref{Eq:EFTNLRen} is valid below the UV-cutoff scale $M_{st}$, only axionic wormholes with radius $a(0)$ bounded from below by $M_{st}^{-1}$ are to be considered. The Giddings-Strominger wormholes~\cite{Giddings:1987cg} associated to $\xi$ with radius $a_{GS}(0) = 0.14 w_\xi^{1/2} f_\xi^{-1/2} M_{Pl}^{-1/2}$ are thus supported by charges $w_\xi > 48 f_\xi M_{Pl} M_{st}^{-2} \sim {\cal O}(10^{6})$, such that the scalar potential only acquires extremely suppressed corrections from these gravitational instantons with action ${\cal S}_{GS} \sim w_\xi\, M_{Pl}/f_{\xi}$ and the inflationary model is not compromised by the considerations in~\cite{Montero:2015ofa}. Also the Abbott-Wise wormholes~\cite{Abbott:1989jw} with radius $a_{AW} (0) =0.077 w_\eta^{2/3} \lambda^{1/6} M_{Pl}^{-1}$ can only be supported by large $w_\eta$-charge ($w_\eta > 47 M_{Pl}^{3/2} M_{st}^{-3/2} \lambda^{-1/4} \sim {\cal O}(10^{5})$) and a sub-Planckian decay constant $f$, such that their contribution to the energy-momentum will dominate over the Giddings-Strominger contribution. Integrating out the Abbott-Wise wormholes then induces $U(1)$ violating corrections:
\begin{equation}
{\cal L}_{\rm WH} = \sum_n \alpha_n  M_{st}^{4-n} \left(\frac{M_{st}}{M_{Pl}}\right)^4   \Phi^n  e^{-{\cal S}_{\rm AW}} + h.c., 
\end{equation}
which equally receive a huge suppression due to the large instanton action ${\cal S}_{\rm AW} \approx w_\eta^{4/3} \lambda^{1/3} \gg 1$ and $M_{st} \ll M_{Pl}$.

The WGC~\cite{ArkaniHamed:2006dz} offers another well-argued criterion to test UV-compatibility with quantum gravity for a field theory. The massless axions in action (\ref{Eq:FullStringModelEFTMS}) provide the required states to satisfy the WGC in the UV for the $U(1)$ symmetry. Below $M_{st}$ the infrared theory retains the global $U(1)$ remnant, which is broken explicitly by non-perturbative effects such as instantons and Euclidean wormholes. Applying the zero-form formulation of the WGC for the axionic inflaton $\xi$ requires us to identify instantons to which the axion $\xi$ couples with decay constant $f_2$ and with an action $S_{inst}$ constrained by $ S_{inst} f_2 \leq M_{Pl}$. Euclidean D-brane instantons~\cite{Blumenhagen:2006xt} form the natural candidates, provided that their intersections with the D-branes supporting the gauge groups in (\ref{Eq:FullStringModelEFTMS}) give rise to a superpotential of the form (\ref{Eq:NonPertSuper}), with the amplitude $A$ now a function of the chiral fermions. In practice, the axionic direction $(p_1,p_2)$ coupling to this E-brane instanton will not align with the direction coupling to the non-Abelian gauge group, as in Fig.~\ref{Fig:AxionModuliSpace}, such that the resulting effective decay constant $f_2$ will be sub-Planckian and the WGC is satisfied by such an E-brane instanton. Moreover, if the cycle volume of the E-brane is larger than the cycle volume of the non-Abelian gauge group, the 't Hooft operator (\ref{Eq:GaugeInstantonMassTerm}) will be the dominant non-perturbative effect~\cite{CPSS2} and the WGC-loophole for axions can be realized~\cite{Brown:2015iha}.

\section{Conclusions}
In this Letter, we consider the most generic EFT for closed string axions arising from Type II compactifications with D-branes and work out how strong non-Abelian dynamics
generate masses for the scalar excitations (infladrons and axions). Our set-up allows the realization of ``natural-like inflation" with a firm control on quantum corrections to the scalar potential. The infladron back-reaction leads to a flattening of the axion scalar potential, which can alleviate the tension with the observational CMB data. Moreover, gravitational corrections such as axionic wormholes are argued to be suppressed in this scenario. The proper functioning of this scenario relies on the identification of a non-supersymmetric de Sitter vacuum through moduli stabilization~\cite{CPSS2}, in which the $U(N)$ D-brane stack is wrapped around a small internal cycle~\cite{Svrcek:2006yi,Honecker:2017air}. The promising aspects of our model invite further investigation into quantum gravity constraints for strongly coupled chiral gauge theories coupled to axions and potential constraints upon confrontation with more refined formulations of the WGC~\cite{ArkaniHamed:2006dz, Brown:2015iha,Hebecker:2015rya,Heidenreich:2015wga}, to map out the contours of the swampland~\cite{Ooguri:2006in}.

\subsection{Acknowledgments}

We thank Kiwoon Choi, Arthur Hebecker, Aitor Landete, Fernando Marchesano, Liam McAllister, Miguel Montero, Francisco G.~Pedro, Angel Uranga and Clemens Wieck for useful discussions and suggestions.  G.S.~is supported in part by the DOE grant DE-SC0017647
and the Kellett Award of the University of Wisconsin.
He would also like to thank the University of Chinese Academy of Sciences for hospitality while part of this work was carried out. 
W.S.~is supported by the ERC Advanced Grant SPLE under contract ERC-2012-ADG-20120216-320421, by the grants FPA2015-65480-P (MINECO/FEDER EU) and IJCI-2015-24908 from the MINECO, and the grant SEV-2012-0249 of the ``Centro de Excelencia Severo Ochoa" Programme.


\begin{thebibliography}{99}
\bibitem{comment0} 
For simplicity we will restrict ourselves to models with generation-independent $U(1)$ charges, i.e. $q_L^i=q_L$ and $q_R^i=q_R$ for $i = 1, \ldots, N_f$.
\bibitem{Comment1} 
This condition considers only the one-loop effects due to massless gauge bosons and fermions, while their respective superpartners are assumed to be massive with  a sufficiently high mass scale.
\bibitem{Comment2} 
If the constraint $N_f < \frac{34  N_c^3}{13 N_c^2 - 3}$ is satisfied, the Yang-Mills theory does not exhibit a Banks-Zaks fixed point in the IR.
\bibitem{Comment3}
Strongly coupled N-JL interactions are characterized by a non-trivial Nambu-Goldstone vacuum configuration breaking the chiral $U(1)$ symmetry without the need of strongly coupled non-Abelian physics. The implications of the N-JL-mechanism in light of inflationary models is discussed elsewhere~\cite{CPSS2}.
\end{thebibliography}
\end{document}